\documentclass[twocolumn,amsmath]{revtex4}

\usepackage{amsfonts}
\usepackage{amsmath}
\usepackage{amssymb}
\usepackage{graphicx}%
\begin{document}

\title{Neutrino mixing matrix  in terms of neutrino mass matrix \\and its Frobenius covariants}
\author{M. I. Krivoruchenko}
\affiliation{National Research Centre ''Kurchatov Institute'', Pl. Akademika Kurchatova 1, 123182 Moscow, Russia}
\author{F. \v Simkovic}
\address{Faculty of Mathematics, Physics and Informatics$\mathrm{,}$ Comenius University in Bratislava, 842 48 Bratislava, Slovakia}
\address{Institute of Experimental and Applied Physics, Czech Technical University in Prague, 110 00 Prague, Czech Republic}

\begin{abstract}
An explicit expression is derived for the mixing matrix of Majorana neutrinos in terms of the mass matrix and its Frobenius covariants.
Illustrative scenarios are taken as examples where the suggested formula is used to obtain the mixing matrix from the mass matrix.
\end{abstract}

\maketitle

\section{Introduction}
\renewcommand{\theequation}{I.\arabic{equation}}
\setcounter{equation}{0}

In the search for generalizations of the Standard Model (SM), a special place is given to neutrinos, as the lightest, electrically neutral fermions. The mass matrix of neutrinos contains about a third of all the free SM parameters; it may contain important information about ways to generalize SM.

The presence of simple algebraic structures is helpful in the theoretical analysis of neutrino-related processes. Algebraic methods in neutrino physics are quite popular (see, e.g., \cite{Bellandi:1997,Aquino:1999,Fogli:1999,Denton:2020,Denton:2022}).
We mention \cite{Denton:2022} for an interesting theorem on the relationship of the modulus of matrix elements of the mixing matrix with the eigenvalues of the mass matrix and its minors. Frobenius covariants \cite {Roger:1991} have also found a natural application in the physics of massive and mixed neutrinos \cite{Fogli:1999,Krivoruchenko:2023}.

The aim of this paper is to derive an explicit analytical formula for the mixing matrix in terms of the neutrino mass matrix and its Frobenius covariants. The techniques we developed enable the analytical recovery of both the modules and the phases of the mixing matrix elements.

\section{Majorana neutrino mass matrix}
\renewcommand{\theequation}{II.\arabic{equation}}
\setcounter{equation}{0}


Majorana neutrinos are described by a Lagrangian
\begin{equation} \label{LM}
\mathcal{L}_{M}=\frac{1}{2}\bar{\nu}_{\alpha} i\hat{\nabla}\nu _{\alpha}-\frac{1}{2}%
\bar{\nu}_{\alpha} \left( M_{\alpha \beta}+i\gamma _{5}N_{\alpha \beta}\right) \nu _{\beta},
\end{equation}
whrere $\alpha, \beta = 1,..,n$, and $n$ is the number of flavors. We work with the real four-component spinors:
$\nu _{\alpha} = \nu _{\alpha}^c \equiv \nu _{\alpha}^*$.
The definition of gamma matrices is the same as in the monograph \cite{Bjorken:1964},
with the exception that the Majorana representation with imaginary $\gamma$-matrices is employed instead of the standard one.
The left and right neutrinos are defined by $\nu _{\alpha L} = \frac{1}{2}(1 - \gamma_5) \nu _{\alpha}$
and $\nu _{\alpha  R} = \frac{1}{2}(1 + \gamma_5) \nu _{\alpha}$, respectively.
In what follows, the components of the vectors of flavor space are denoted by the letters of the Greek alphabet $\alpha, \beta, \ldots$,
the components of the vectors diagonalizing the mass matrix are denoted by the letters of the Latin alphabet $r, s, \ldots$.

The conditions $\bar{\nu}_{\alpha }\nu _{\beta}=\bar{\nu}_{\beta}\nu _{\alpha }$
and $\bar{\nu}_{\alpha}i\gamma _{5}\nu _{\beta}=\bar{\nu}_{\beta}i\gamma _{5}\nu_{\alpha }$
imply that the mass matrices $\mathbb{M} = \{M_{\alpha \beta}\}$ and $\mathbb{N} = \{N_{\alpha \beta}\}$ are symmetric.
The hermiticity of the Lagrangian (\ref{LM}) guarantees the hermiticity of $\mathbb{M}$ and $\mathbb{N}$.
They are actually real because they are symmetric, and so
$\mathbb{M} = \mathbb{M}^* = \mathbb{M}^T$ and
$\mathbb{N} = \mathbb{N}^* \,= \mathbb{N}^T$.
We define a $4n \times 4n$ real matrix:
\begin{equation} \label{matrix12}
\hat{\mathbb{M}} = \mathbb{M} +i\gamma _{5}\mathbb{N},
\end{equation}
which obeys $\hat{\mathbb{M}}^* = \hat{\mathbb{M}}$, and two $n\times n$  complex symmetric matrices
\begin{equation} \label{matrix}
\mathbb{M}_{\pm }= \mathbb{M} \pm i\mathbb{N}.
\end{equation}
One can verify that
$\gamma_0 \hat{\mathbb{M}}^\dagger \gamma_0 =  \hat{\mathbb{M}}$
and $\mathbb{M}_{\pm}^\dagger = \mathbb{M}_{\mp}$.
We also use notations $\mathbb{M}_{R} = \mathbb{M}_{+}$ and $\mathbb{M}_{L} = \mathbb{M}_{-}$.

The propagator of neutrinos
\begin{eqnarray}
\hat{\mathbb{S}}_F(p) = \frac{1}{\hat{p} - \hat{\mathbb{M}}} \label{SF}
\end{eqnarray}
is determined by the quadratic form of the Lagrangian.
The matrix $\hat{p} - \hat{\mathbb{M}}$ has the dimension $4n$. The energy eigenvalues
are grouped by pairs $\pm \sqrt{m_i^2 + \mathbf{p}^2}$, where the masses $m_i$ are assumed to be positive ($i = 1,..,n$).

The chirality-state projections take the form
\begin{eqnarray}
\hat{\mathbb{S}}_F(p) \frac{1 \pm \gamma_5}{2} = (\hat{p} + \mathbb{M}_{\mp}) \frac{1}{p^2
- \mathbb{M}_{\pm}\mathbb{M}_{\mp}}\frac{1 \pm \gamma_5}{2}.
\end{eqnarray}
The propagator poles, which correspond with the zeros of the characteristic polynomial
\begin{eqnarray} \label{character}
p(\lambda) = \det||\lambda - \mathbb{M}_{\pm}\mathbb{M}_{\mp}|| = 0,
\end{eqnarray}
determine the mass spectrum.
There are normalized vectors $|r + \rangle$ which obey
\begin{equation}  \label{eigen}
\mathbb{M}_{+}\mathbb{M}_{-} | r + \rangle = \lambda_r |r + \rangle.
\end{equation}
The eigenvalues $\lambda_r = m_r^2$ are real positive:
\begin{equation}
\lambda_r = \langle r + | (\mathbb{M}_{-})^\dagger \mathbb{M}_{-} |r + \rangle > 0.
\end{equation}
Given that $|r + \rangle$ is the normalized eigenvector of $\mathbb{M}_{+}\mathbb{M}_{-}$,
\begin{equation} \label{bases}
|r - \rangle \equiv \frac{1}{m_r} \mathbb{M}_{-} |r + \rangle
\end{equation}
is the normalized eigenvector of $\mathbb{M}_{-} \mathbb{M}_{+}$ with the same eigenvalue $\lambda_r$,
and
\begin{equation} \label{bases2}
|r + \rangle = \frac{1}{m_r} \mathbb{M}_{+} |r - \rangle.
\end{equation}

The eigenvector coordinates can be considered as complex conjugate in the basis of flavor states:
\begin{equation}  \label{symm}
\langle \alpha |r + \rangle ^* = \langle \alpha |r - \rangle.
\end{equation}

Equation (\ref{bases}) is equivalent to the condition
\begin{equation} \label{symm5}
\langle r - |\mathbb{M}_{L}|s + \rangle = m_r \delta_{rs},
\end{equation}
which shows that for $n$ flavors, $\mathbb{M}_{L}$ can  be diagonalized by two unitary matrices:
\begin{equation} \label{VMU}
\mathbb{V}^{\dagger}\mathbb{M}_{L}\mathbb{U} = \mathrm{diag}(m_1,\ldots,m_n),
\end{equation}
with
\begin{eqnarray}
V_{\alpha r} &=& \langle \alpha|r - \rangle, \\
U_{\alpha r} &=& \langle \alpha|r + \rangle.
\end{eqnarray}
By virtue of Eq.~(\ref{symm}), $\mathbb{V}$ is expressed in terms of $\mathbb{U}$: $\mathbb{V} = \mathbb{U}^*$,
so that
\begin{equation}  \label{symm2}
\mathbb{U}^{T}\mathbb{M}_{L}\mathbb{U} = \mathbb{V}^{T}\mathbb{M}_{R}\mathbb{V} = \mathrm{diag}(m_1,\ldots,m_n).
\end{equation}
The matrix $\mathbb{U}$ is known as the Pontecorvo--Maki--Nakagawa--Sakata (PMNS) mixing matrix.

The neutrino states are projected onto the mass eigenstates
with the help of the Frobenius covariants. In the multiplicative representation, the Frobenius covariants
of the neutrino mass matrix take the form
\begin{eqnarray} \label{eq3}
|r \pm \rangle \langle r \pm| \equiv \mathbb{F}_{r \pm} = \prod_{s \neq r}\frac{\mathbb{M}_{\pm}\mathbb{M}_{\mp} - \lambda_s}{\lambda_r - \lambda_s}. \label{expl7}
\end{eqnarray}
We remark that
\begin{equation}  \label{symm4}
\mathbb{F}_{r \pm} = \mathbb{F}_{r \pm}^\dagger = \mathbb{F}_{r \mp}^*  = \mathbb{F}_{r \mp}^T,
\end{equation}
$\mathbb{F}_{r \pm}$ satisfies
\begin{eqnarray}
\mathbb{F}_{r \pm} \mathbb{F}_{s \pm} &=& \mathbb{F}_{r \pm}\delta_{rs}, \\
\mathbb{F}_{r \pm} \mathbb{M}_{\pm} &=& \mathbb{M}_{\pm}\mathbb{F}_{r \mp}, \label{comm} \\
\langle \alpha |\mathbb{F}_{r \pm}|\gamma \rangle &=& \langle \gamma |\mathbb{F}_{r \mp}|  \alpha \rangle, \label{inter} \\
\sum_{\alpha = 1}^n\langle\alpha|\mathbb{F}_{r\pm}|\alpha\rangle &=& 1. \label{norm}
\end{eqnarray}
The first line shows that $\mathbb{F}_{r \pm}$ is a projection operator.
The third line is obtained with the use of
Eqs.~(\ref{symm}), (\ref{eq3}), and (\ref{symm4}).

\section{Neutrino mixing matrix}
\renewcommand{\theequation}{III.\arabic{equation}}
\setcounter{equation}{0}

From the previous section, we need Eqs.~(\ref{symm5}) and (\ref{eq3}), which we write in the form
\begin{eqnarray}
\sum_{\alpha \beta } U_{\alpha r}U_{\beta s}\langle \alpha |\mathbb{M}_{L}|\beta \rangle = m_{r}\delta _{rs}, \label{eq1} \\
U_{\alpha r}U_{\beta r}^* = \langle \alpha |\mathbb{F}_{r +}|\beta \rangle. \label{eq2}
\end{eqnarray}
The second line immediately determines the modulus of matrix elements
\begin{equation}
\left\vert U_{\alpha r}\right\vert ^{2}=\langle \alpha |\mathbb{F}%
_{r+}|\alpha \rangle  > 0.
\end{equation}

Let us define the phase factors in the matrix elements:
\[
U_{\alpha r}=e^{i\varphi _{\alpha r}}|U_{\alpha r}|=e^{i\varphi _{\alpha r}}%
\sqrt{\langle \alpha |\mathbb{F}_{r+}|\alpha \rangle }.
\]%
Equation (\ref{eq2}) gives
\begin{equation}
U_{\alpha r}=\frac{1}{U_{\gamma r}^*}\langle \alpha |\mathbb{F}%
_{r+}|\gamma \rangle =e^{i\varphi _{\gamma r}}\frac{\langle \alpha |\mathbb{F%
}_{r+}|\gamma \rangle }{\sqrt{\langle \gamma |\mathbb{F}_{r+}|\gamma \rangle
}}.  \label{eq33}
\end{equation}
The left part does not depend on $\gamma$.
We substitute this expression for Eq.~(\ref{eq1}), perform the summation over $\beta$, and take the diagonal term $r=s$.
The result is
\begin{eqnarray}
&&e^{2i\varphi _{\gamma r}}\sum_{\alpha }\langle \alpha |\mathbb{F}_{r+}|\gamma \rangle
\langle \alpha |\mathbb{M}_{L}\mathbb{F}_{r+}|\gamma \rangle \nonumber \\
&~& ~~~~~~~~~~~~~= m_{r}\langle \gamma |\mathbb{F}_{r+}|\gamma \rangle.~~
\end{eqnarray}
Using Eqs.~(\ref{comm}) and (\ref{inter}), we perform the summation over $\alpha$ and obtain
\begin{eqnarray}
e^{2i\varphi _{\gamma r}}\langle \gamma |\mathbb{M}_{L}\mathbb{F}%
_{r+}|\gamma \rangle =m_{r}\langle \gamma |\mathbb{F}_{r+}|\gamma \rangle.
\end{eqnarray}
The phase factors are now
\begin{equation}
e^{2i\varphi _{\gamma r}}=\frac{m_{r}\langle \gamma |\mathbb{F}_{r+}|\gamma
\rangle }{\langle \gamma |\mathbb{M}_{L}\mathbb{F}_{r+}|\gamma \rangle }=
\frac{\langle \gamma |r+\mathbb{\rangle }}{\langle \gamma |r-\mathbb{\rangle
}}
\end{equation}
and
\begin{equation} \label{final}
U_{\gamma r}=V_{\gamma r}^*=\sqrt{\frac{m_{r}}{\langle \gamma |\mathbb{M}_{L}\mathbb{F}%
_{r+}|\gamma \rangle }}\langle \gamma |\mathbb{F}_{r+}|\gamma \rangle .
\end{equation}
The neutrino mixing matrix is explicitly represented in this equation
in terms of the neutrino mass matrix and its Frobenius covariants as determined
by the flavor basis.

Given Eq. (\ref{final}),
it is possible to randomly choose the signs of the columns in the matrix $\mathbb{U}$
since the root function is defined up to the sign. The uncertainty is due to the
invariance of Eqs. (\ref{eq1}) and (\ref{eq2}) under the transformation $U_{\alpha r} \rightarrow U_{\alpha r}^{\prime}=U_{\alpha
r}\epsilon_{r}$ with $\epsilon_r = \pm $1.
Because $\mathbb{U}$ is a unitary matrix, the matrix $\mathbb{U}^{\prime}$ is likewise unitary.
To unambiguously determine $\mathbb{U}$, it is necessary to fix an arbitrary value of $\gamma=1,\ldots,n$. The signs of the square root $\sqrt{\langle\gamma|\mathbb{M}_{L}\mathbb{F}_{r+}|\gamma\rangle}$ are then chosen
for each $r=1,\ldots,n$. The equation
\begin{align}
U_{\alpha r}=\sqrt{\frac{\langle\gamma|\mathbb{M}_{R}\mathbb{F}_{r-}|\gamma\rangle
}{m_{r}}}
\frac{\langle\alpha|\mathbb{F}_{r+}|\gamma\rangle}{\langle
\gamma|\mathbb{F}_{r+}|\gamma\rangle} \label{final2}
\end{align}
is used to determine each individual matrix element of the mixing 
matrix.

Equation (\ref{final2}) is singular in $\langle\gamma|\mathbb{F}_{r+}|\gamma\rangle$.
If $\langle\gamma|\mathbb{F}_{r+}|\gamma\rangle = 0$, we substitute $\gamma \rightarrow (\gamma \mod(n)) + 1$
as many times as necessary to obtain a result different from zero.
Due to normalization (\ref{norm}), there exists a $\gamma$ such that $\langle\gamma|\mathbb{F}_{r+}|\gamma\rangle \neq 0$.

\vspace{3mm}
\textbf{Example 1}.
Let's take a look at an example of neutrino mass matrix, considered in Ref. \cite{Krivoruchenko:2023}:
\begin{equation}
\hat{\mathbb{M}}=\lambda^{3}+i\gamma_{5}\lambda^{6},
\end{equation}
where $\lambda^{a}$ are the Gell-Mann matrices.
The eigenvectors of $\mathbb{M}_{R}\mathbb{M}_{L}$ can easily be found using the MAPLE symbolic computation
software package \cite{MAPLE}:
\begin{equation}
\tilde{U}_{\alpha i} = \left[
\begin{array}{ccc}
1 & 0 & 0 \\
0 & -i/g & ig \\
0 & 1 & 1%
\end{array}%
\right] ,  \label{example8}
\end{equation}%
where $g=(1+\sqrt{5})/2$. The unnormalized eigenvectors are represented by the columns of $\tilde{\mathbb{U}}$.
The normalization restores the mixing matrix
\begin{equation} \label{ex1}
U_{\alpha r} =
\left( \sum_{\beta } |\tilde{U}_{\beta r}|^2 \right)^{-1/2}\tilde{U}_{\alpha r}.
\end{equation}
The calculation based on
Eq.~(\ref{final2}) reproduces the mixing matrix (\ref{ex1}) numerically for positive signs of the square root of $\langle\gamma|\mathbb{M}_{R}\mathbb{F}_{r-}|\gamma\rangle$
with any initial value of $\gamma$.

\vspace{3mm}
\textbf{Example 2}.
Let us consider a more realistic, so-called tri-bimaximal mixing model of Majorana neutrinos with the mass matrix \cite{Bilenky:2018}
\begin{equation}
\mathbb{M}_{L} = \left[
\begin{array}{ccc}
x & y & y \\
y & x+v & y-v \\
y & y-v & x+v
\end{array}
\right],  \label{example8}
\end{equation}
where
\begin{eqnarray}
x &=& \frac{2}{3}m_1 + \frac{1}{3}m_2, \\
y &=& - \frac{1}{3}m_1 + \frac{1}{3}m_2, \\
v &=& - \frac{1}{2}m_1 + \frac{1}{2}m_3.
\end{eqnarray}
The Frobenius covariants can be found to be
\begin{eqnarray}
\mathbb{F}_{1\pm} &=& \left[ \begin {array}{rrr} 2/3&-1/3&-1/3\\ -1/3&1/6&1/6\\ -1/3&1/6&1/6\end {array} \right], \\
\mathbb{F}_{2\pm} &=& \left[ \begin {array}{rrr} 1/3& 1/3& 1/3\\ ~~ 1/3&1/3&1/3\\ 1/3& ~~~ 1/3& ~~ 1/3\end {array} \right], \\
\mathbb{F}_{3\pm} &=& \left[ \begin {array}{rrr} ~~~~~ 0  &   0&   0\\    0&1/2& -1/2 \\   0& \,-1/2& 1/2\end {array} \right].
\end{eqnarray}

These matrices obey Eqs.~(\ref{symm4}) - (\ref{norm}).
The precise correspondence of $\mathbb{U}$ with the standard definition \cite{Bilenky:2018} is obtained by choosing a positive sign
for the square root of $\langle\gamma|\mathbb{M}_{L}\mathbb{F}_{r+}|\gamma\rangle$ at $r=1,2$ and a negative sign at $r=3$:
\begin{eqnarray}
\mathbb{U} = \left[ \begin {array}{rrr}
2/\sqrt{6}  &1/\sqrt {3} & 0\\
-1/\sqrt {6}&1/\sqrt {3} &-1/\sqrt {2} \\
-1/\sqrt {6}&1/\sqrt {3} & 1/\sqrt {2}
\end {array} \right].
\end{eqnarray}

Other neutrino mixing matrix scenarios are also discussed (see \cite{Razzaghi:2022jgq,Ismael:2022emm} and references therein).

\section{Conclusion}
\renewcommand{\theequation}{V.\arabic{equation}}
\setcounter{equation}{0}

It was previously shown that the amplitudes of the main processes associated with neutrinos can be determined in terms of the mass matrix explicitly without involving the mixing matrix. Given that the mixing matrix is expressed analytically in terms of the neutrino mass matrix and its Frobenius covariants, this paper generalized these conclusions to all processes associated with neutrinos.

The main result of this paper is the formula (\ref{final2}) for the mixing matrix in terms of the neutrino mass matrix and
its Frobenius covariants defined in the flavor basis.
This result is more general than the eigenvector-eigenvalue theorem on the relationship between the modules of the elements of the mixing matrix and the eigenvalues of the mass matrix and its diagonal minors.
Equation (\ref{final2}) establishes analytically
both the modules and the phases of the mixing matrix elements.

Fitting the experimental observables in neutrino-involving processes
with respect to the mass matrix makes it possible to determine the mass matrix with better accuracy compared to methods using
the mixing matrix at intermediate stages.

\acknowledgements
F.\v{S}. acknowledges support by the VEGA Grant Agency of the Slovak
Republic under Contract No. 1/0607/20 and by the grant Large Research
Infrastructures (LRI) under Contract LM2 023063.

\end{document}